\documentclass[twocolumn,
showpacs,
preprintnumbers,
amsmath,
amssymb,
groupedaddress,
superscriptaddress
]{revtex4}

\usepackage{graphicx}
\usepackage{dcolumn}
\usepackage{bm}
\usepackage{xcolor}
\usepackage{verbatim}

\usepackage{hyperref}
\usepackage[mathlines]{lineno}

\usepackage{epstopdf}

\begin{document}


\title{Examining the event-shape dependent modifications to charged-particle\\
   transverse momentum spectra and elliptic flow in p-Pb collisions at\\
   energies available at the CERN Large Hadron Collider}

\author{Somnath Kar}
\email{somnathkar11@gmail.com}
\affiliation{Key Laboratory of Quark and Lepton Physics (MOE) and Institute of Particle Physics, Central China Normal University, Wuhan 430079, China}

\author{Subikash Choudhury}
\email{subikash@fudan.edu.cn}
\affiliation{Key Laboratory of Nuclear Physics and Ion-beam Application (MOE) and Institute of Modern Physics, Fudan University, Shanghai 200433, China}

\author{Xiaoming Zhang}
\email{xiaoming.zhang@mail.ccnu.edu.cn}
\affiliation{Key Laboratory of Quark and Lepton Physics (MOE) and Institute of Particle Physics, Central China Normal University, Wuhan 430079, China}

\author{Daicui Zhou}
\email{dczhou@mail.ccnu.edu.cn}
\affiliation{Key Laboratory of Quark and Lepton Physics (MOE) and Institute of Particle Physics, Central China Normal University, Wuhan 430079, China}

\date{\today}

\begin{abstract}
Purported signatures of collective dynamics in small systems like proton-proton (pp) or proton-nucleus (p-A) collisions still lack unambiguous understanding. Despite the qualitative and/or quantitative agreement of the data to hydrodynamic models, it has remained unclear whether the harmonic flows in small systems relate to the common physical picture of hydrodynamic collectivity driven by the initial geometry. In the present work, we aim to address this issue by invoking a novel concept of Event Shape Engineering (ESE), which has been leveraged to get some control of the initial geometry in high-energy heavy-ion collisions. We utilise ESE by constructing a reference flow vector, $q_{2}$ that allows to characterise an event based on it's ellipticity. Applying this technique on a data set, simulated from a 3+1D viscous hydrodynamic model EPOS3, we study the event-shape dependent modifications to some of the bulk properties like, inclusive transverse momentum ($p_{T}$) spectra and $p_{T}$-differential $v_{2}$ for p-Pb collisions at 5.02 TeV. Selecting events on the basis of different magnitudes of reference flow vector $q_{2}$, we observe a hint of event-shape induced modifications of $v_{2}$ as a function of $p_{T}$ but, the inclusive $p_{T}$-spectra of charged particles seem to be insensitive to this event-shape selection. 
\end{abstract}


\maketitle

\section{Introduction}
Hydrodynamic modelling has remained the most successful description to the
properties of the bulk matter produced in the collisions of heavy nuclei at
ultra-relativistic energies~\cite{intro_1, intro_2}. The efficacy of hydrodynamic calculations
have not only allowed to characterize the medium produced in these collisions
as a strongly interacting fluid, but also, presented unambiguous evidences
that relate final state momentum space azimuthal anisotropies to the initial spatial inhomogeneities.
It is generally perceived that an inviscid 
hydrodynamic evolution efficiently translates these
inhomogenities in the initial state to the final state momentum space azimuthal
anisotropies-quantified by the coefficients $v_{n}$s in the Fourier decomposition of
the azimuthal distributions of produced particles in a plane transverse to the beam
axis~\cite{intro_5, intro_6, intro_7}.

Since long, the applicability of the hydrodynamic models were thought to be limited
to large and extended systems like the one produced in heavy-ion collisions.
However, only recently, it was realized that the dynamical behaviour of the medium
produced in hadron-hadron or hadron-on-ion collisions (small systems) exhibit
remarkable similarity to  those of the heavy-ions~\cite{intro_8, intro_9, intro_10,
intro_11}. Notably, the agreement of hydrodynamic calculation to unexpectedly large values of anisotropic flow coefficients
triggered speculations whether the collisions of small systems are also dominated by strong final state interactions.
~\cite{intro_12, intro_13}. However, it
must be mentioned that the strongly interacting nature of the medium produced in
large systems were not only inferred from the agreement of hydrodynamic calculations to 
the measurements of $p_{\rm T}$-differential yields and anisotropic flow
coefficients at low-$p_{\rm T}$, but also, corroborated by the concurrent
observations of the energy loss of high-$p_{\rm T}$ particles/jets which by-far
remain elusive in small systems~\cite{intro_14, intro_14a, intro_14b}. In addition,
the so called hallmark of the hydrodynamic collectivity, in particular, the sizeable
magnitudes of flow harmonics in small systems are also confronted by distinctly
different suite of physical interpretations where strong final state interactions
have not been invoked~\cite{intro_16, intro_17, intro_18, intro_19,
intro_20}. This counter-intuitive observation of the hydro-like collectivity, in the
absence of the jet-quenching, therefore, underscores the importance of studying the
emergent phenomenon of collective dynamics in small systems with all forms of
available tools at our disposal.

Recently, a test of hydrodynamization in small systems was conducted at RHIC
with shape engineered collision species; p-Au, d-Au and He-Au collisions, producing
intrinsically circular, elliptic and triangular configurations respectively, in
their initial geometry~\cite{intro_21}. It was argued that the imprints of this
initial geometry will be reflected at the final stage provided the hydrodynamic
collectivity prevails. For example, if the system has an intrinsic elliptic or
triangular shape, hydrodynamic collectivity would favor an ordering between the
final state elliptic ($v_{2}$) and triangular ($v_{3}$) flow
coefficients~\cite{intro_13, intro_21N}. The measurements of $v_{2,3}$ by the PHENIX
collaboration indeed presented some evidence in favor of this conjectured
correlations between the initial geometry and hydro-expected ordering in the flow
patterns~\cite{intro_21}. Therefore, further experimental investigations
on such initial geometry dependent ordering of harmonic flows at higher 
$\sqrt{\rm s}$ might be timely and desirable to corroborate the claims of common hydrodynamic 
paradigm across widely different system sizes
~\cite{intro_23}. However, till date, the scopes of 
exploring the fluid dynamical picture in small systems with intrinsically different
initial geometries at the LHC energies are unlikely. Notwithstanding this
limitation, the influence of initial geometry on the final state momentum space
anisotropy of the produced particles can therefore be examined with an alternative 
novel technique namely, the Event Shape Engineering (ESE)~\cite{intro_24}. 

In the framework of Glauber-like initial condition followed by the hydrodynamic 
evolution, the event-by-event fluctuations in the distributions of the initial
nuclear matter is manifested as large spread in the distributions of initial and 
final state anisotropies~\cite{intro_25}. This can be eventually exploited to 
further categorize events into different classes of initial geometry but at 
comparable multiplicity. This technique of selecting events on the basis of initial 
geometry is generally referred to as the Event Shape Engineering. 
A key component of this technique is the determination of reference flow vectors $q_{\rm n}$s (n = 2, 3
etc.) in the momentum space, which by construction are correlated to n$^{\rm th}$ order harmonic (for n $<$ 4)
flow coefficients and hence to the corresponding orders of asymmetries at the initial co-ordinate space~\cite{intro_26}.
Here it must be mentioned, unlike the hydrodynamic descriptions that relate the flow harmonics to 
the initial geometry, the flow-like signals in the Color Glass Condensate (CGC) 
Effective Field Theory (EFT) theory, on the other hand, are attributed to initial 
state gluon momentum correlations which depend on a saturation length scale 
(1/Q$_{\rm s}$) via event multiplicity.  Therefore, the flow harmonics within the 
CGC theory are supposedly independent of the event geometry. As a result, the ESE 
technique could be used as an effective tool to distinguish the underlying origin of
harmonic flows in small systems.

Since the original proposal, the ESE technique has been applied to several 
experimental measurements either to constrain the flow-induced backgrounds or to 
investigate the degree of correlations between different orders of flow 
harmonics~\cite{intro_27, intro_28, intro_29}. In this work, we examine the response
of the bulk properties of the produced medium at the final state to the variations in
the magnitudes of the initial spatial asymmetries by applying the ESE technique to a
small system like p-Pb collisions at 5.02 TeV. Using an event-by-event 3+1D viscous
hydrodynamic model, EPOS3, we investigate the modifications to the inclusive
yields and the elliptic flow coefficient, $v_{2}$ of charged particles as a function
of $p_{\rm T}$, for an ensemble of events with higher or lower than the average bulk
elliptic flow anisotropy, quantified by the reduced second order flow-vector, $q_{\rm
2}$.\\
Remainder of this paper is structured as follows. In section II we provide a brief
account of the hydrodynamic model EPOS3, followed by the analysis details in section
III. In section IV we present the results and finally we discuss and summarize in
section V.
\section{EPOS3: The Model}
EPOS3 is built on a $p$QCD inspired framework for Gribov-Regge multiple parton scattering approach, where an individual scattering generates a longitudinally stretched colored flux-tube (strings) with transverse kinks carrying $p_{\rm T}$ from the initial hard scatterings~\cite{EPOS1}. These flux-tubes eventually break into pairs of string segments that lead to the production of final state particles following Schwinger mechanism of string fragmentation.

A high multiplicity event in EPOS3  is characterized by a highly dense medium of colored strings produced from a large number of parton-parton interactions. Under such condition, several strings overlap to each
other which prevent them to hadronize independently, as described above. 
In this situation, EPOS3  classifies these strings to constitute
either jets or the bulk matter. Based on an energy loss formalism, fate of the strings are decided i.e, whether they will 
be a part of the bulk matter or emerge out as high-$p_{\rm T}$ particles/jets~\cite{EPOS2}. 
If the fractional energy loss of string segments exceed a certain threshold which is a model dependent parameter,
they constitute the bulk matter, the so-called core, that undergo a viscous hydrodynamic expansion and hadronize by the usual
Cooper-Frye formalism at a hadronization temperature, T$_{\rm H}$. Rest of the string segments form corona and hadronize by the usual Schwinger mechanism.
In general, EPOS3 is able to describe some aspects of the data in small collision systems reasonably well.
This includes the double-ridge structure in the two-particle angular correlations, the $p_{\rm T}$-dependence and the
characteristic mass ordering of  $v_{\rm 2}$, among others~\cite{EPOS4, EPOS5}.
\begin{figure}[htb!]
\centering
\includegraphics[scale=0.70,keepaspectratio]{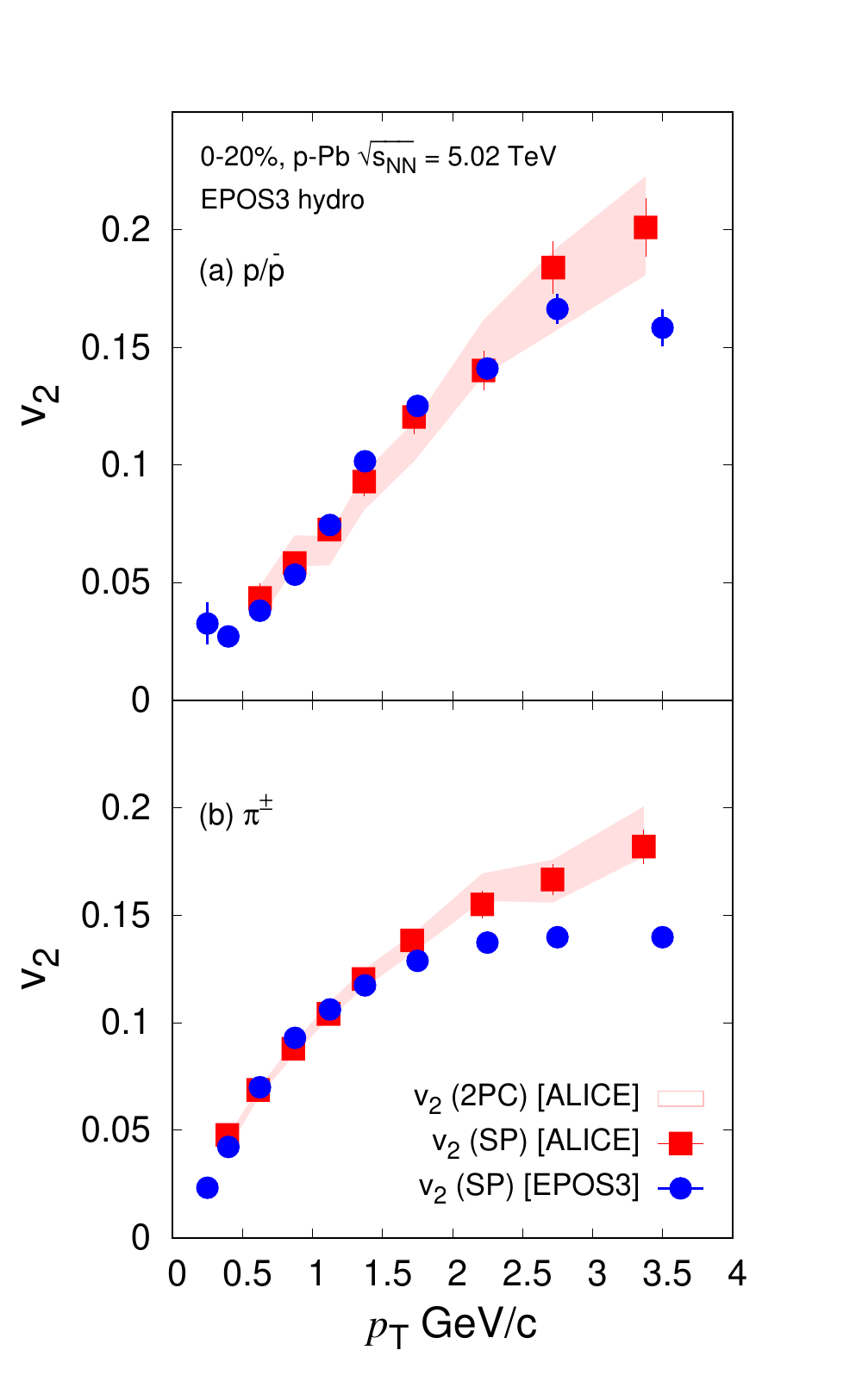}	
\caption{[Color online] Elliptic flow parameter $v_{2}$ for (a) protons ($p + \bar{p}$) and (b) pions ($\pi^{+} + \pi^{-}$) as a function of $p_{ \rm T}$ calculated from full hydrodynamic simulation of  EPOS3 for 0-20\% most central p-Pb events at $\sqrt{\rm s_{\rm NN}}$ = 5.02 TeV.}
\label{v2pt}
\end{figure}

To verify, whether the simulated EPOS3 event-samples can emulate some aspects of the p-Pb data, we calculate 
$v_{\rm 2} ( p_{\rm T} )$ of pions and protons for 0-20\% most central p-Pb events and compare the same with published ALICE results~\cite{EPOS6} in Fig.~\ref{v2pt}. In one of our previous publications~\cite{EPOS2}, we also compared the multiplicity dependent invariant yields of identified particles as a function $p_{\rm T}$ calculated from the EPOS3-generated events to the data. In both cases agreement with the data are well-founded.
Having observed a good agreement between data and simulated events, we proceed further to testify the central theme of our present work.

\section{Analysis}
\subsection{ Centrality and the event-shape ($q_{\rm 2}$) determination}
EPOS3 generated event samples are first sub-sampled into multiplicity (centrality) classes based on the particle multiplicity in the pseudorapidity coverage, 2.8 $ < \eta < $ 5.1, corresponding to ALICE V0A detector acceptance~\cite{ana1}. Details of the centrality selection from the minimum bias EPOS3 generated p-Pb samples can be found here~\cite{EPOS2}.\\

In a given multiplicity interval, these events are further categorized into different classes of reduced second-order harmonic flow vector,
$q_{2}$ defined as~\cite{intro_26, ana3}:

\begin{equation}
 q_{2} = |Q_{2}| / \sqrt{M} 
\end{equation}
where M corresponds to number of particles used in the calculation of the second-order harmonic flow vector, $Q_{2}$. The definition for the flow vector $Q_{2}$ is
\begin{equation}
 |Q_{2}| = \sqrt{ Q_{2x}^{2} +  Q_{2y}^{2}}, 
\end{equation}
where $Q_{2x}, Q_{2y}$ correspond to the cosine and sine component of flow vector $Q_{2}$ respectively.

\begin{figure}[htb!]
\centering
\includegraphics[scale=0.55,keepaspectratio]{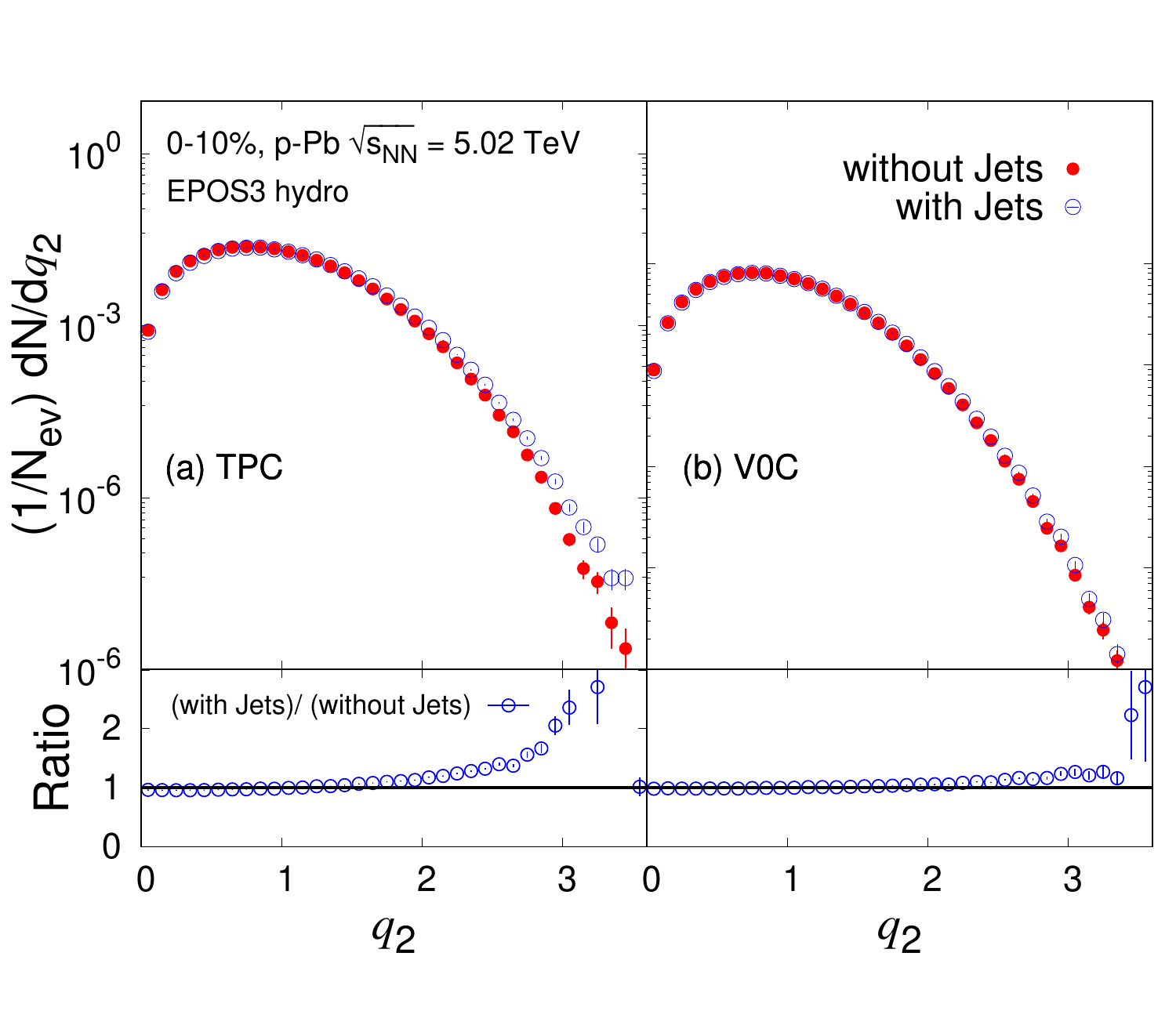}
\caption{[Color online] Distributions of second order reference flow vector $q_{ \rm 2}$ calculated in the equivalent $\eta$-acceptance of (a) ALICE-TPC and (b) ALICE-V0C with and without subtractions of jetty events. The lower panels show the ratio of $q_{2}$ distributions with and without subtraction of jetty events for both $q_{\rm 2}^{\rm TPC}$ and $q_{\rm 2}^{\rm V0C}$.}
\label{q2TPCV0C}
\end{figure}

In this work, we calculate $q_{\rm 2}$ in the $p_{\rm T}$-range 0.2 $ < p_{\rm T} < $ 20 GeV/c at two different pseudo-rapidity region; one in the mid-rapidity, $| \eta |< $  0.3
and other in the forward rapidity, -1.7 $ < \eta < $  -3.7. Former has an overlap with the detector 
coverage of ALICE-TPC~\cite{ana4} while, later is equivalent to ALICE-V0C~\cite{ana5} acceptance. Hereafter in the text and in figures, 
we will refer $q_{\rm 2}$ calculated in these two regions
as $q_{\rm 2}^{\rm TPC}$ and $q_{\rm 2}^{\rm V0C}$ respectively. Figure \ref{q2TPCV0C} shows the $q_{\rm 2}^{\rm TPC}$ ($q_{\rm 2}^{\rm V0C}$) distributions
for 0-10\% highest multiplicity events. 

As the particle production in small systems are dominated by the $p$QCD processes, flow vectors so obtained are presumably vulnerable to large non-flow effect from dijets, di-minijets and also resonance decays. In general, the contributions from the non-flow effects scale inversely with particle multiplicity M, where M  could be the number of particles used in the determination of flow vector, $Q_{\rm 2}$. Therefore, we will mostly focus on the high-multiplicity
events rendering an automatic reduction to the non-flow effects. 
Although, such a choice can naturally reduce the non-flow contributions in larger systems where particle multiplicity is originally high 
but this may not be strictly true for small systems where the overall particle multiplicity is less. Thus, to further mitigate the 
non-flow related contributions to $q_{\rm 2}$ we invoke rejection of events that has a jet of minimum jet-$p_{\rm T}$ (without background subtraction) of 5 GeV/c. To do so we make use of the jet reconstruction technique where, jets are reconstructed using the standard anti-$k_{\rm T}$ jet-finding algorithm in the \textsc{Fastjet} package~\cite{ana6, ana7} for resolution parameter R = 0.2. \\
The effect of removing jetty events can be readily observed from the ratios of $q_{\rm 2}^{\rm TPC}$ distributions, before
and after the removal of jetty events in Fig.~\ref{q2TPCV0C}. Towards the higher values of $q_{\rm 2}^{\rm TPC}$, ratios differ from unity by 
50\% or more, implying a substantial jet-bias. But the difference is less prominent for $q_{\rm 2}^{\rm V0C}$, suggestive of its robustness against jet contamination. 
However, the observed effect for $q_{\rm 2}^{\rm V0C}$ may be completely model dependent.
A possible reason that we can think-of is the drop in dijet or di-minijet yields in EPOS3, away from the mid-rapidity.
However, $q_{\rm 2}^{\rm V0C}$ has an advantage over $q_{\rm 2}^{\rm TPC}$ as it provides a large natural pseudorapidity ($|\Delta\eta|$) separation
between regions of calculating $Q$-vectors and the observables of physics-interest (which is calculated here within $0.5 < |\eta| <$ 1). 
This is rather crucial for the removal of the auto-correlations and the correlated now-flow effects. 
But, for $q_{\rm 2}^{\rm TPC}$ we could only afford a maximum $|\Delta\eta|$ gap of 0.2 unit, because of our choice of limited
$\eta$ coverage of $\pm$1, to be able to comply with ALICE-TPC acceptance.

\begin{figure}[htbp]
\centering
\includegraphics[scale=0.36,keepaspectratio]{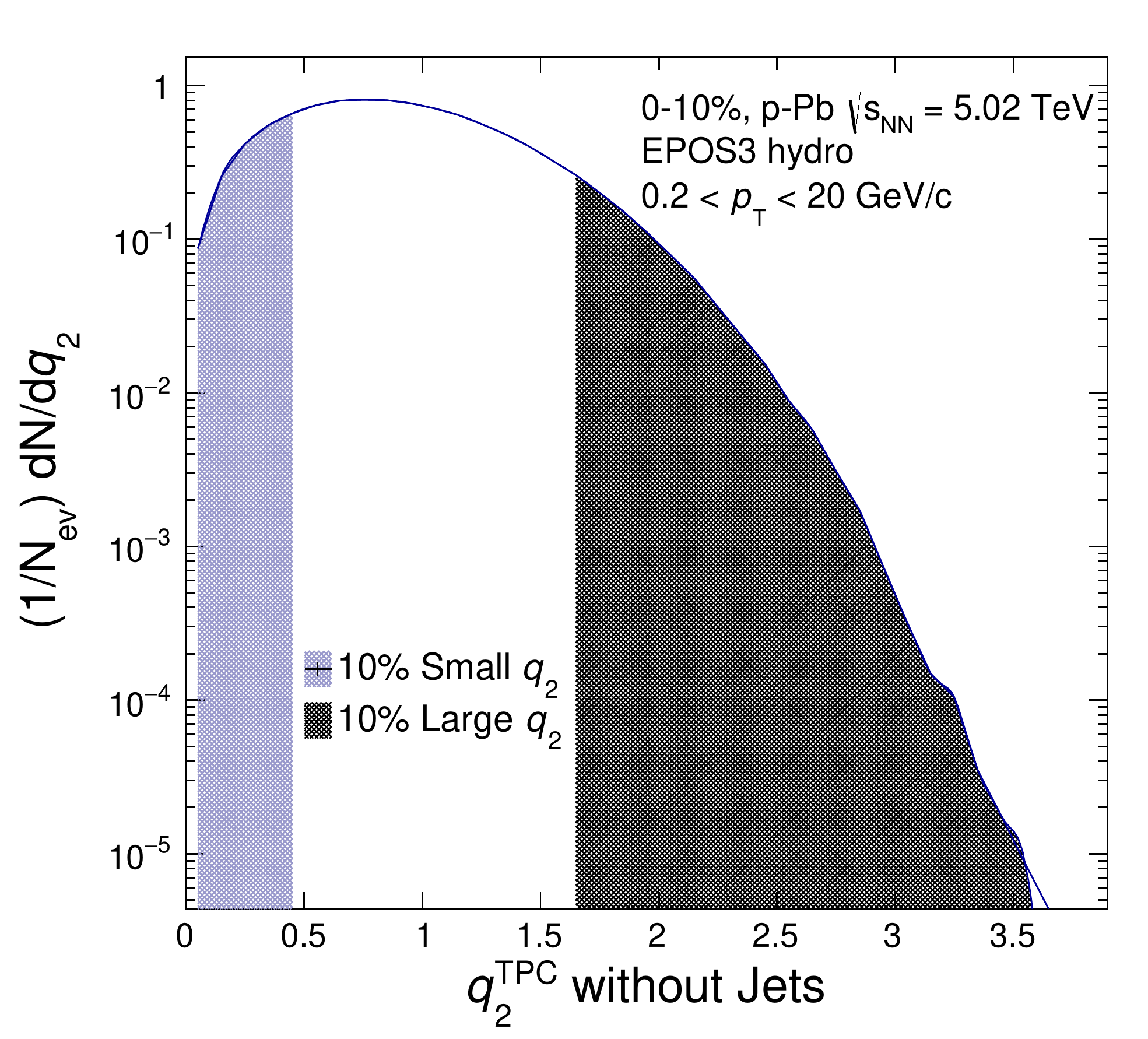}
\caption{[Color online] Representations of 10\%-large (small) selection areas in the $q_{2}$ distribution calculated in the TPC region after the removal of jetty events. }
\label{q2TPCSub}
\end{figure}
Figure~\ref{q2TPCSub} shows the jet subtracted $q_{\rm 2}^{\rm TPC}$ distribution for 0-10\% highest multiplicity
events and the shaded regions in the same correspond to top and bottom 10\% of events with highest and lowest values 
of $q_{\rm 2}^{\rm TPC}$ respectively .
We will calculate the physics observables in this highest (0-10\%) and lowest (90-100\%) 
10\% bins of $q_{\rm 2}^{\rm TPC}$ as well as $q_{\rm 2}^{\rm V0C}$, which will be referred in remainder of the text as large and small $q_{\rm 2}^{\rm TPC} $or $q_{\rm 2}^{\rm V0C}$ respectively. 
To be mentioned, because of limited statistics, we report our results averaged over an interval of 10\% 
multiplicity bin, but the ESE-selection classes are defined based on $q_{\rm 2}$ percentiles obtained from 1\% multiplicity bin-width 
in-order to avoid any trivial fluctuations in $q_{\rm 2}$ due to fluctuations in the particle multiplicity.
\section{Results}
\subsection{Transverse momentum distributions}
\begin{figure*}[hbpt]
\centering
\includegraphics[scale=0.56,keepaspectratio]{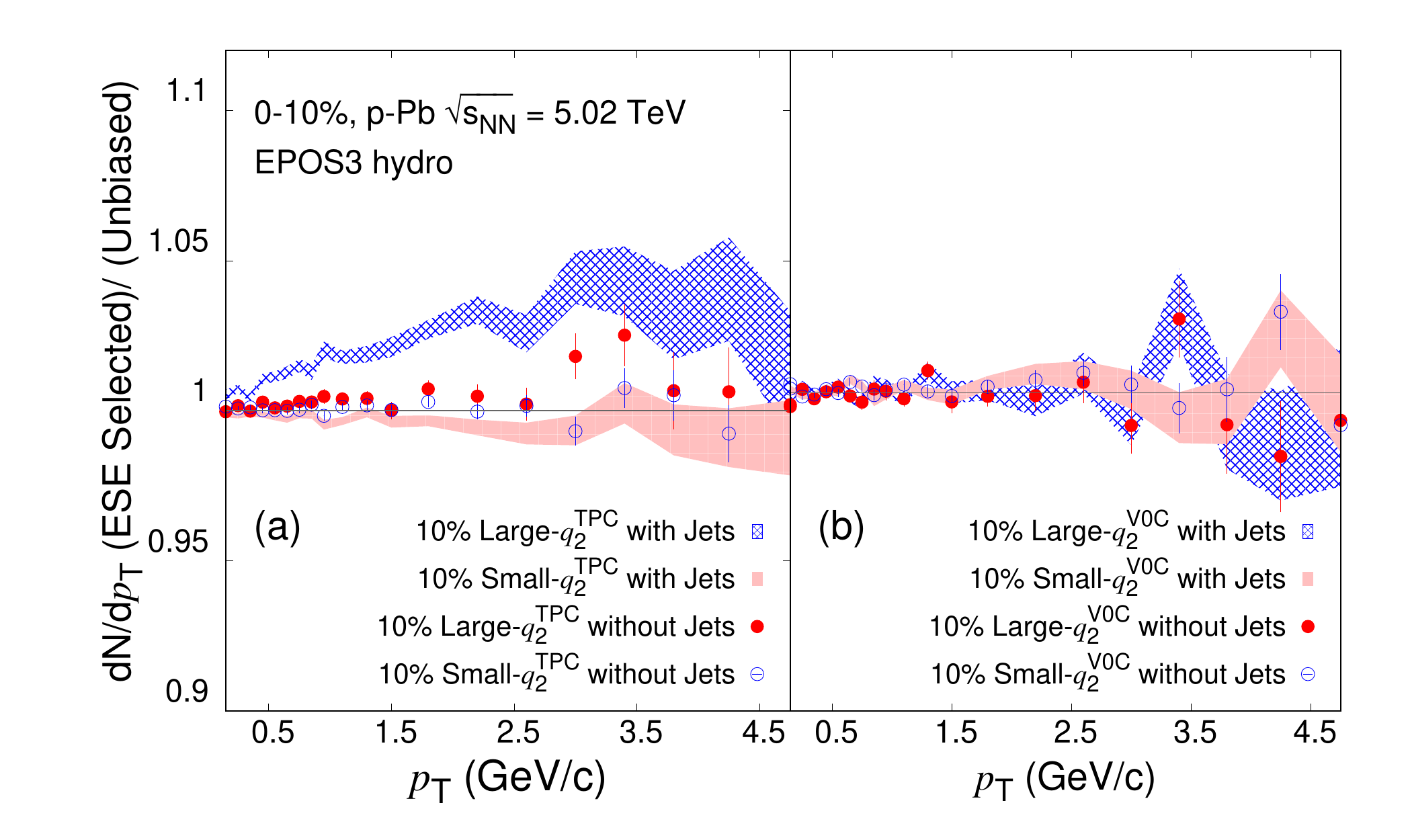}
\caption{[Color online] Ratio of charge particle yields in ESE-selected events w.r.t unbiased sample as a function of transverse momentum for $q_{2}^{\rm TPC}$ (a) and $q_{2}^{\rm V0C}$ (b) for the EPOS3 simulated events with hydro in p-Pb collisions at $\sqrt{\rm s_{NN}} =$ 5.02 TeV. A comparison of with and without removal of jetty events is also shown for both the regions of $q_{2}$ selections.}
\label{q2ESE_Yield_withWOJets}
\end{figure*}

Effect of event-shape selection is first studied on the single inclusive charged particle $p_{\rm T}$ spectra
for large and small-$q_{\rm 2}^{\rm TPC}$ and $q_{\rm 2}^{\rm V0C}$ event samples and reported in 
Fig.~\ref{q2ESE_Yield_withWOJets}.  
As mentioned, to avoid overlap with the $\eta$ range of $q_{\rm 2}^{\rm TPC}$ ($|\eta| < $ 0.3),
the $p_{\rm T}$ distributions of unidentified charged particles are calculated in the range 0.5  $< |\eta| < $ 1.0.
In order to study how the jet-contamination in $q_{\rm 2}$ affects the event-shape selections and hence the modifications of $p_{\rm T}$ spectra in shape-engineered event samples,
we calculate the ratios of charged particle $p_{\rm T}$-spectra in shape-biased to shape-unbiased events on the basis of
10\% highest and lowest $q_{\rm 2}$ percentiles, derived from $q_{\rm 2}$ distributions with and without jet contaminations as shown in 
Fig.~\ref{q2TPCV0C}. Blue and red bands in Fig.~\ref{q2ESE_Yield_withWOJets} correspond to the results obtained from
event-shape selection based on the $q_{\rm 2}$ distributions including jet-bias. On the other hand, markers in Fig.~\ref{q2ESE_Yield_withWOJets}
represent the same results except the $q_{\rm 2}$ percentiles are determined from $q_{\rm 2}$-distributions without jet-bias. The effect
of jet-contamination in $q_{\rm 2}$ is manifestly evident from the comparison of these two cases. When the $q_{\rm 2}$ percentiles are 
extracted from the $q_{\rm 2}$ distributions including jetty events, ratios of $p_{\rm T}$-spectra in shape-biased to unbiased event
samples exhibit an increasing trend with increasing $p_{\rm T}$. However, upon removal of the jetty events and re-calculating
the $q_{\rm 2}$ percentiles based on the $q_{\rm 2}$ distribution without jet-contamination, the ratio is rather flat and consistent 
with unity. This suggests that the apparent hardening of the spectral shape,
in particular, in large-$q_{\rm 2}$ events could be because the mean of the $q_{\rm 2}$ distribution is shifted towards the higher values due to systematic bias from the jet-dominated events.

Also for $q_{\rm 2}^{\rm V0C}$ selection, the aforementioned exercise is repeated to study the possible modifications
to the spectral shape in large and small-$q_{\rm 2}^{\rm V0C}$ event samples relative to shape unbiased sample. In a marked contrast
to $q_{\rm 2}^{\rm TPC}$, shape selection on the basis $q_{\rm 2}^{\rm V0C}$ is seemingly unaffected by the jet contamination.
This agrees to our previous observation in Fig.~\ref{q2TPCV0C}, where the impact of removal of jetty events was found to be insignificant on $q_{\rm 2}^{\rm V0C}$ distributions itself.

As we observe that the removal of jetty events has large impact on the shape dependent charge particle yields, 
we therefore proceeded to do some systematic checks to establish robustness of these results. Since, we consider only reconstructed jet-$p_{\rm T}$ without background subtraction, there could be chances of over-estimation of jet-$p_{\rm T}$ resulting in removal of events in excess to what is needed. Therefore, to understand whether our final results are stable against this proposed jetty event removal technique, we repeated the analysis varying the minimum $p_{\rm T}$ of the input particles that are fed into jet-reconstruction algorithm.

\begin{figure*}[htbp]
\centering
\includegraphics[scale=0.70,keepaspectratio]{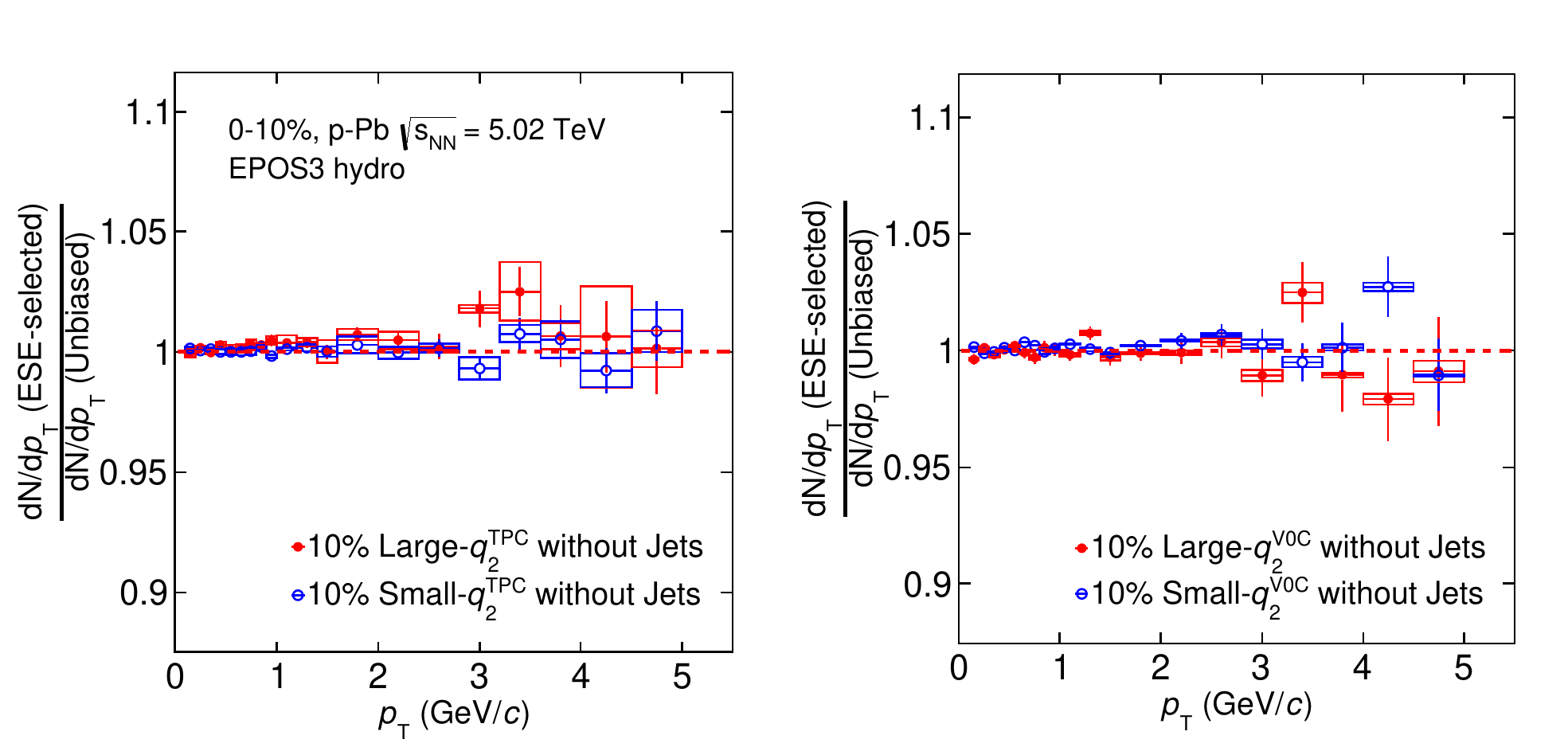}
\caption{[Color online] Systematic variations of ratio of charge particle yields as a function of transverse momentum in ESE-selected events w.r.t unbiased sample after the removal of jetty events for (a) $q_{2}^{\rm TPC}$ and (b) $q_{2}^{\rm V0C}$.}
\label{q2ESE_Yield_TPCV0CwithoutJets}
\end{figure*}

The minimum $p_{\rm T}$  of input particles taken so far as a default choice is 0.3 GeV/c. For systematics, this
value is changed to 0.15  and 0.5 GeV/c respectively. Subsequently, jets are reconstructed with the corresponding sets of input particles,
followed by removal jetty events from the $q_{\rm 2}$ distribution in the same way as already mentioned.
The open-boxes in Fig.~\ref{q2ESE_Yield_TPCV0CwithoutJets} 
represent the systematic variation on the ratios plotted in Fig.~\ref{q2ESE_Yield_withWOJets} and indicated by solid and open markers.
The systematic changes in the ratio are well within the limits of current statistical uncertainties. 

\subsection{ Elliptic flow }

In this section we report the results of elliptic flow coefficient of charge particles in 
unbiased and shaped-engineered
event samples. The elliptic flow coefficient, $v_{\rm 2}$, as a function of $p_{\rm T}$ 
is calculated in the pseudorapidity range 0.5 $< | \eta| < $ 1.0, using the Scalar Product method~\cite{SP1, SP2}.
In this method, an event is divided into sub-events without an overlap in pseudorapidity. This is done by defining
atleast two sub-events separated by an $\eta$-gap. Here we have defined two sub-events A and B covering the
eta range -0.5 $<\eta <$ -1.0 and 0.5 $<\eta <$ 1.0 respectively, and calculated $v_{\rm 2} (p_{\rm T})$
according to the relation
\begin{equation}
v_{2} \{SP\} (p_{T})= \frac{<u_{2, i} Q_{2}^{*}/M>}{\sqrt{<Q_{2, A}Q_{2, B}^{*}/M_{A}M_{B}}>} ,
\end{equation}

where $u_{\rm 2, i} = e^{2\phi_{i}}$ is the unit vector of i$^{\rm th}$ particle of interest, 
$\phi_{i}$ is the corresponding azimuthal angle and $Q^*_{2}/M$ is the multiplicity normalized 2$^{\rm nd}$ order flow vector.
In the denominator,
$Q_{2, \rm A}$ (M$_{A}$) and $Q_{2, \rm B}$ (M$_{B}$) are the second-order flow vectors (multiplicity) in the sub-event A and B, 
respectively.
The angular bracket in the numerator indicates the average over all particles of interest. 
To suppress non-flow contributions to $v_{2}$, the unit flow vector, $u_{2,\rm i}$ and the flow vector $Q_{2}$
are always evaluated from different sub-events.

\begin{figure}[h!]
\centering
\includegraphics[scale=0.32,keepaspectratio]{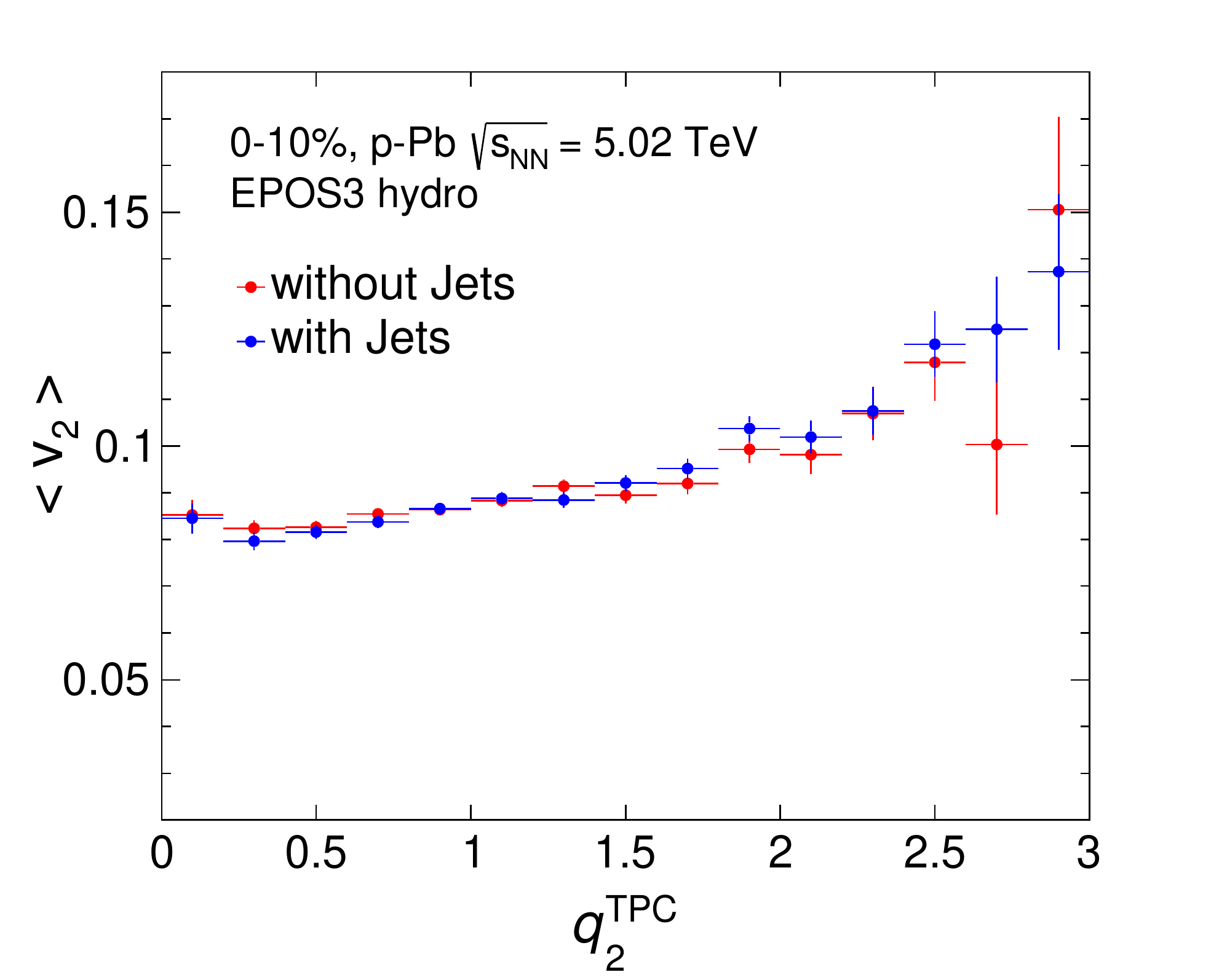}	
\includegraphics[scale=0.32,keepaspectratio]{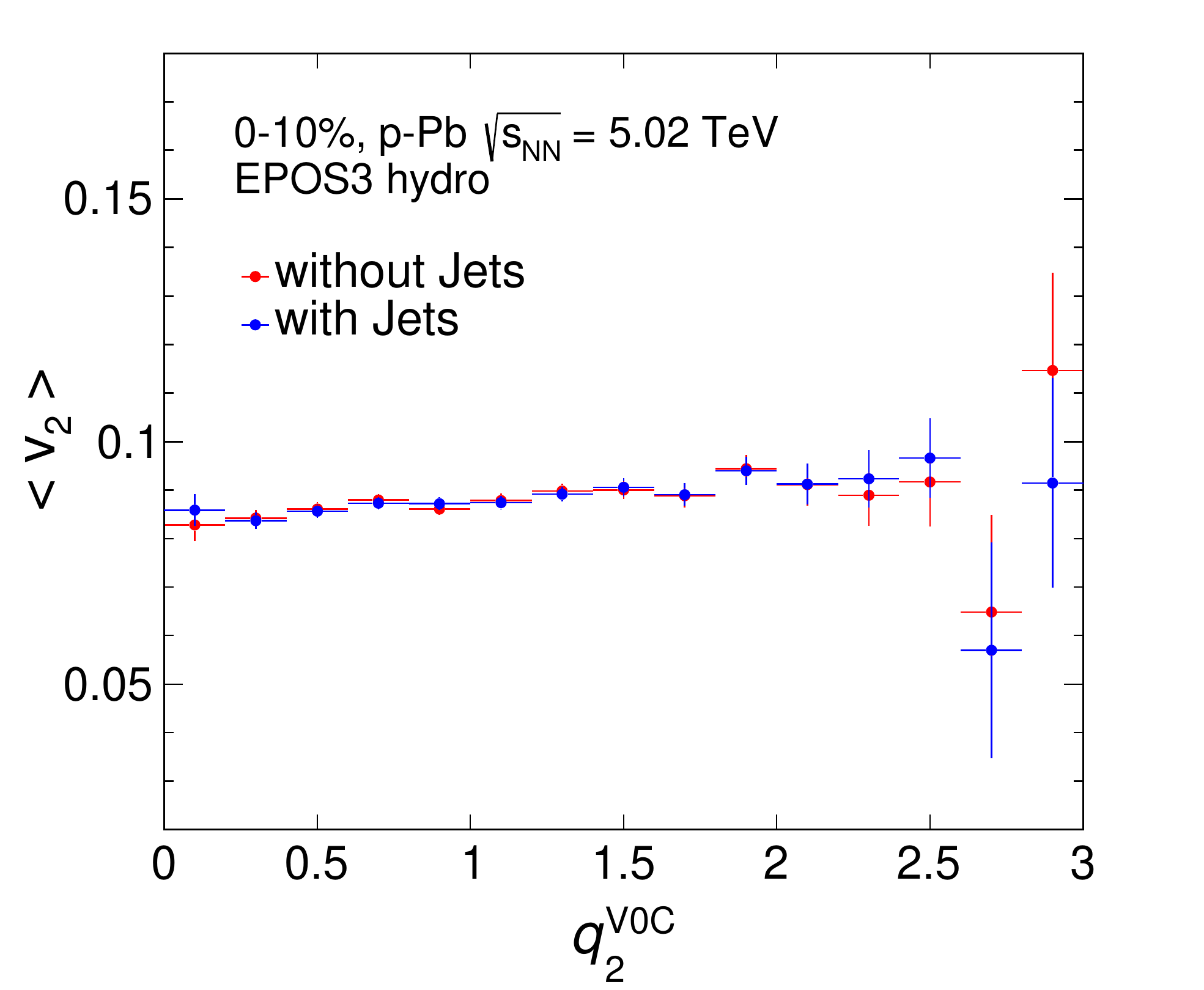}		
\caption{[Color online] Correlations between average elliptic flow coefficient $v_{2}$ and $q_{\rm 2}$ measured in (a) TPC and (b) V0C regions for unidentified charged particles, with and without removal of jetty events from $q_{\rm 2}^{\rm TPC}$ and $q_{\rm 2}^{\rm V0C}$.}
\label{v2_ESE_q2}
\end{figure}

Figure~\ref{v2_ESE_q2} shows the correlation between the $p_{\rm T}$-average elliptic flow coefficient, $<v_{2}>$, and $q_{\rm 2}$
for $q_{\rm 2}^{\rm TPC}$ [Fig~\ref{v2_ESE_q2}(a)] and $q_{\rm 2}^{\rm V0C}$ [Fig~\ref{v2_ESE_q2}(b)]. The $q_{\rm 2}$ values are calculated under two conditions:
with (blue) and without (red) jet contribution. 
The $<v_{2}>$ exhibits a slight increasing trend for both $q_{2}^{\rm TPC}$ and $q_{2}^{\rm V0C}$, but the increase is rather
sharp for $q_{2}^{\rm TPC} >$ 2. This could be due to some correlated residual non-flow effect as the $|\eta|$-gap
available for $q_{2}^{\rm TPC}$ is small.

\begin{figure*}[htbp]	
\centering
\includegraphics[scale=0.60,keepaspectratio]{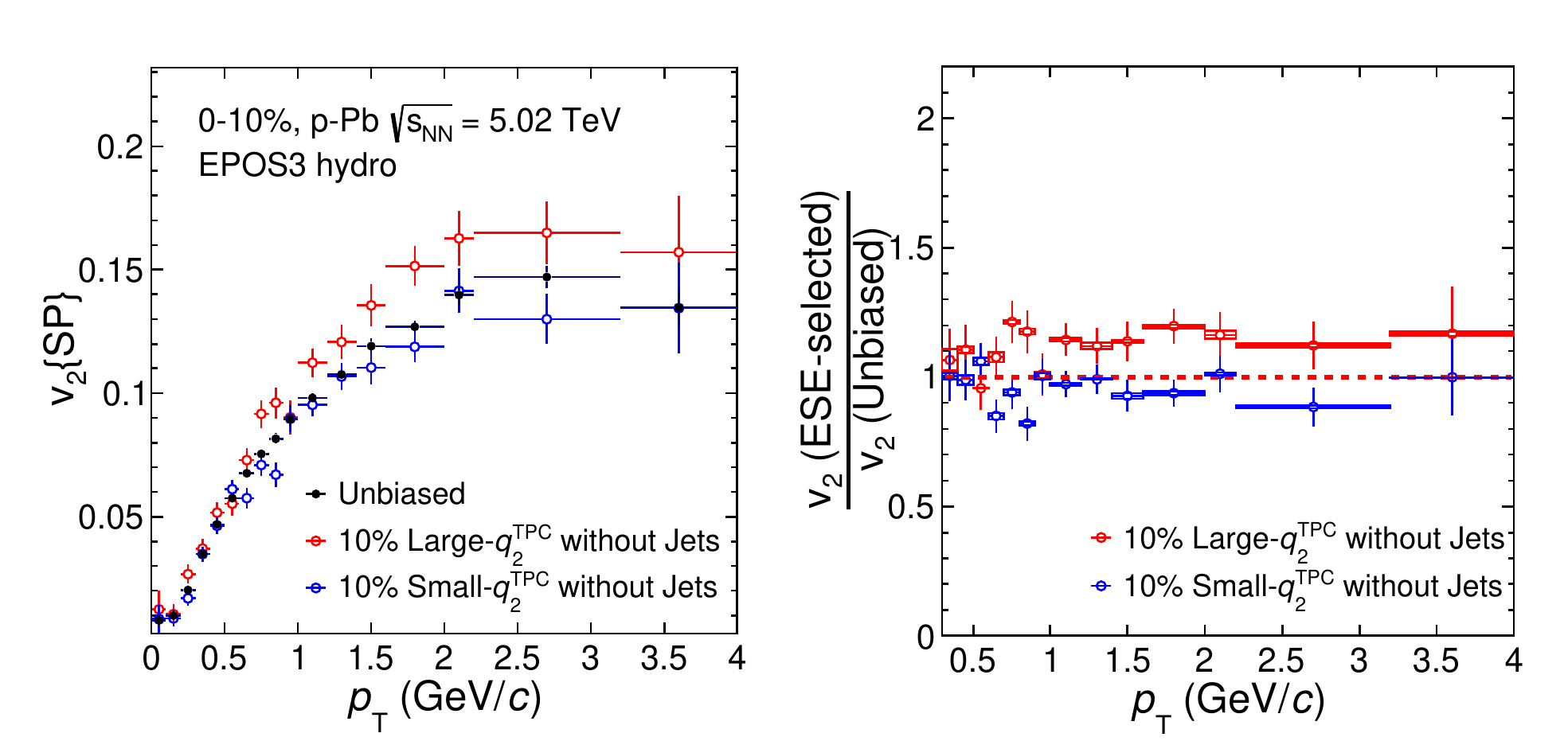}	
\includegraphics[scale=0.60,keepaspectratio]{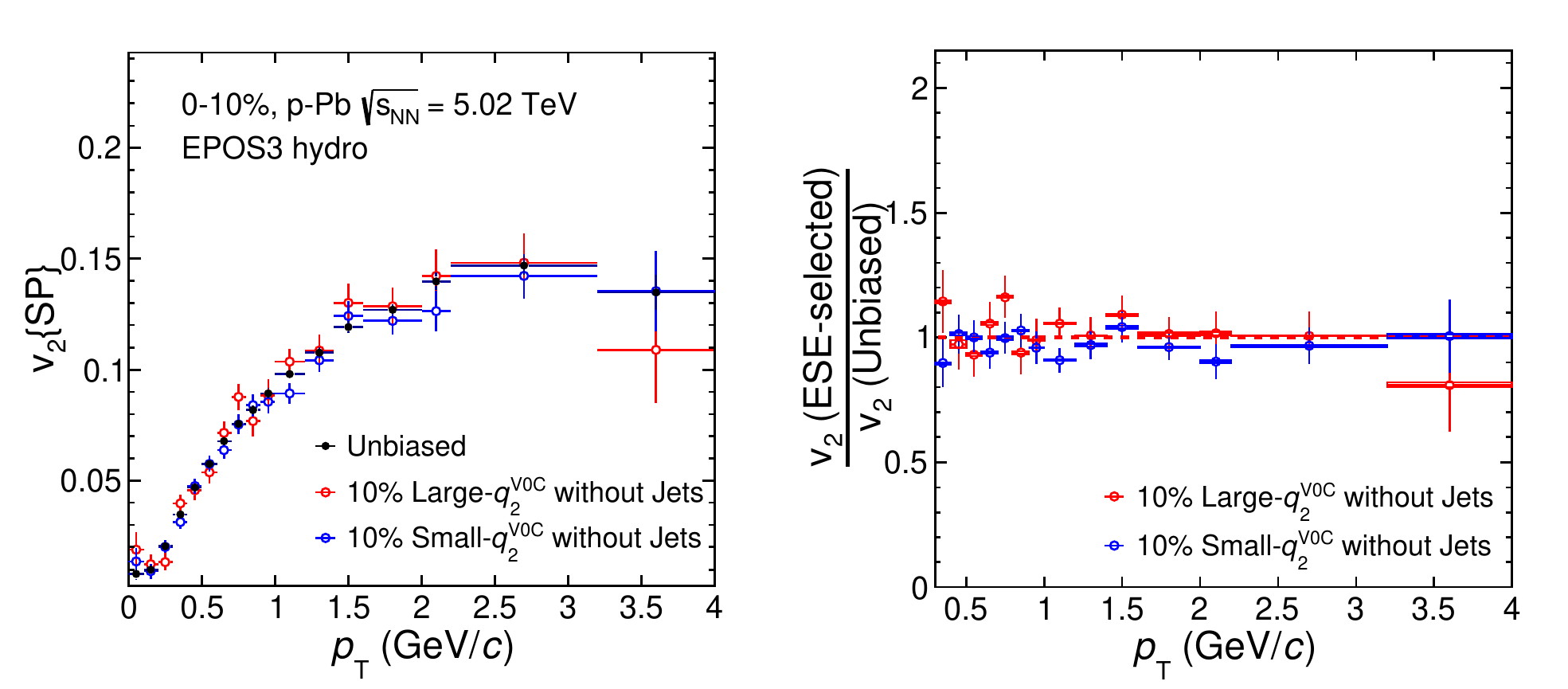}	
\caption{[Color online] Elliptic flow coefficient $v_{2}$ as a function of $p_{\rm T}$ in ESE-selected and unbiased event samples and the ratio of ESE-selected event samples to the unbiased one for $q_{\rm 2}^{\rm TPC}$ (a \& b) and $q_{\rm 2}^{\rm V0C}$ (c \& d) after the removal of jetty events. The systematic variations to the ratios of $v_{2}$ for the ESE-selected event samples to the unbiased sample are shown with open boxes which are however too small to see with naked eyes.}
\label{v2_ESE_TPCV0CJetsub}
\end{figure*}

Figure~\ref{v2_ESE_TPCV0CJetsub} shows $v_{2}$ as a function of $p_{\rm T}$ in large, small and unbiased-$q_{2}$ 
event samples after the subtraction of jetty events from both $q_{\rm 2}^{\rm TPC}$ and $q_{\rm 2}^{\rm V0C}$ . 
The top row of Fig.~\ref{v2_ESE_TPCV0CJetsub} shows the
charged particle $v_{\rm 2} (p_{\rm T})$ in large and small-$q_{2}^{\rm TPC}$ event samples [Fig.~\ref{v2_ESE_TPCV0CJetsub}(a)] 
and the ratios of $v_{\rm 2} (p_{\rm T})$ [Fig.~\ref{v2_ESE_TPCV0CJetsub}(b)] in large and small-$q_{2}^{\rm TPC}$
event samples relative to the shape-inclusive one for the event-shape selection based on $q_{\rm 2}^{\rm TPC}$. 
The same for $q_{\rm 2}^{\rm V0C}$ are shown in the bottom panel [Fig.~\ref{v2_ESE_TPCV0CJetsub}(c \& d)] . 
It can be observed that for 10\% large(small)-$q_{\rm 2}^{\rm TPC}$ selection, $v_{\rm 2} (p_{\rm T})$ changes by 20\% (10\%) 
with no significant $p_{\rm T}$ dependence.
In contrary, no noticeable difference is observed when the event-shape selection is based
on $q_{\rm 2}^{\rm V0C}$. 
We also repeat the same systematic study for $v_{2}$($p_{\rm T})$, as it was done for $p_{ \rm T}$-differential yields in the previous section. 

\section{Discussions and summary}
The role of initial geometry as an essential ingredient to the dynamics of multi-particle angular correlations in relativistic heavy-ion collisions has been established in the light of hydrodynamic calculations that predict strong linear correlations between coefficients of final state azimuthal anisotropy ($v_{n}$, n $<$ 3) and the corresponding initial spatial asymmetry ($\epsilon_{2}$, $\epsilon_{3}$). Off late, studies on small collision systems have also presented evidence that are typical of the standard picture of the hydrodynamic evolution in heavy-ion collisions. Although, generalization of hydrodynamic calculations to small systems has become a standard practise nonetheless, its applicability has remained highly debated. In view of this existing ambiguity on the issue whether the observed features of azimuthal correlations in small systems are consequences of strong final state interactions resulting in hydrodynamic evolution or manifestations of other physical processes related to the initial state gluon correlations, we employ ESE as a tool to probe the degree of correlation between initial geometrical inhomogeneity and final state azimuthal anisotropy. Making use of ESE technique we study modifications to the charged particle transverse momentum spectra and elliptic flow coefficients in shape engineered in 0-10\% central p-Pb events at 5.02 TeV using a 3+1D viscous hydrodynamic model, EPOS3.

Events are first categorized according to the magnitudes of $q_{2}$ vector calculated at different $|\eta|$-acceptances referred to as $q_{2}^{\rm TPC}$ and $q_{2}^{\rm V0C}$. As the determination of $q_{2}$ vectors in small systems are susceptible to non-flow effects from dijets and di-minijets, we eliminate events with jet-$p_{\rm T} > $ 5 GeV/c. The effect of removing jetty events can be immediately observed from Fig.~\ref{q2TPCV0C}. At large values of $q_{2}$  ($>$2) a  surge in the ratio of $q_{2}$  distribution with and without removal of jetty events can be noticed. This could be due to the fact that very large values of $q_{2}$ arise from the events dominated by jet-like processes. 

The ratio of $p_{\rm T}$-differential yields of charged particle spectra in ESE-selected events to those unbiased events shown in Fig.~\ref{q2ESE_Yield_withWOJets} exhibits hardening (softening) in large-$q_{2}^{\rm TPC}$ (small-$q_{2}^{\rm TPC}$) samples when classification was done on the basis $q_{2}^{\rm TPC}$ calculated without removing jetty events. On removal of jet contamination and reclassification of large- and small-$q_{2}^{\rm TPC}$ event samples, no significant difference in the ratios
$p_{\rm T}$-differential yields are observed for $q_{2}^{\rm TPC}$ event-shape selection rather, the ratios of yields in shape biased to unbiased event samples are seen to be consistent with unity. This confirms $q_{2}$ distributions in small systems, in particular, has large non-flow bias.
Similar calculation repeated on the basis of $q_{2}^{\rm V0C}$ are also consistent with unity and shows no effect of jet subtraction. 

\begin{table}[b]
\caption{\label{tab:table1}%
Parameters for bast-wave fit
}
\begin{ruledtabular}
\begin{tabular}{lccr}
\multicolumn{1}{c} ~ & ~
\textrm{Temperature (T$_{kin}$) in GeV} & ~~\textrm{$\beta$} \\
\colrule
Large-${q_{2}}$ & 0.114  & 0.534\\
Small-${q_{2}}$ & 0.115 & 0.531\\
\end{tabular}
\end{ruledtabular}
\end{table}
At this point we recollect, the measurements of event-shape dependent modifications to $p_{\rm T}$ spectra in heavy-ion collisions by ALICE 
\cite{intro_26} revealed, the $p_{\rm T}$ spectra in large and small $q\rm_{2}$ events exhibit significant hardening and softening respectively. 
This has been attributed to the correlation between the event eccentricity and the radial boost i.e events with larger eccentricity have
increased radial push. But with full hydrodynamic simulation of a small system, like the one studied here, we find no such evidence of correlation 
between eccentricity and radial boost. This may be because the initial energy deposition profile
in small systems are so smeared that the average energy-density
and initial eccentricity is either uncorrelated or weakly-correlated. We substantiate on this assertion by extracting kinetic freezeout temperature
T$_{kin}$ and radial boost parameter $\beta$ in large and small-$q_{2}$ event samples via a simultaneous blast-wave fit~\cite{BGBW} to pion, 
kaon and proton $p_{\rm T}$ spectra. The values obtained, tabulated in table-\ref{tab:table1}, suggest in the collisions of small systems radial boost 
or freezeout temperature are either independent or insensitive to the initial event geometry.

Furthermore, we investigate the effect event-shape engineering on both $p_{T}$-differential and $p_{T}$-integrated elliptic flow coefficients, $v_{2}$ at mid-rapidity. Figure~\ref{v2_ESE_q2}, shows an increasing trend in $p_{T}$-average $v_{2}$ for both $q_{2}^{\rm TPC}$ and $q_{2}^{\rm V0C}$ but the increase is more prominent for $q_{2}^{\rm TPC}$ than $q_{2}^{\rm V0C}$. This is most likely because of the reduced sensitivity of $q_{2}^{\rm V0C}$ to the global event-shape together with the longitudinal decorrelation effect which is expected to be large in asymmetric small collision systems. Whereas, for $q_{2}^{\rm TPC}$, we do observe a relatively sharp rising trend of $<v_{2}>$ but we can not completely ignore correlated non-flow effects as the available $\eta$-gap is much less. Similar arguments are also valid for $p_{T}$-differential $v_{2}$ (shown in Fig.~\ref{v2_ESE_TPCV0CJetsub}) which shows sensitivity of event-shape selection largely depend on the choice of $q_{2}$-vector. 

To Summarize, in this article we make an attempt to asses, whether the final state momentum space anisotropies in small systems originate
from correlations limited to few particles or can be linked to global event properties those associated with event-shapes or profile.
In addition, we also realize that the variable used to gauge the event-shape i.e. $q_{2}$ is very much affected by  non-flow components mostly stemming from dijets and di-minijets.
Therefore, we adopt a scheme to minimize non-flow effects by discarding events dominated by jets. Within the current level of uncertainties  we observe event-shape dependent modification of $v_{2}$ are in line with ESE-expectation provided the reference flow vector ($q_{2}$) and  particles of interest are not widely separated in $\eta$.

Experimental verification of this new set of results are certainly warranted in-order to advance our understandings of the initial
conditions and the subsequent spatio-temporal evolutions in so-called small collision systems at relativistic energies.


\section*{Acknowledgements} 
The authors are thankful to Dr. Klaus Werner for providing them with the EPOS3 code and related discussions. This work is supported by the National Natural Science Foundation of China (Grant No. 11805079 \& 11775097), national key research and development program (Grant No. 2018YFE0104700 \& 2015CB856905) and the Grant CCNU18ZDPY04.  




%
%
\end{document}